\documentclass[twocolumn]{aastex62}

\usepackage{epsf}
\usepackage{amsmath}                
\usepackage{amsfonts}               
\usepackage{amssymb}                
\usepackage{epsfig}                 
\usepackage{float}
\usepackage{color}
\usepackage{tabularx}
\usepackage{comment}
\usepackage{makecell}
\usepackage{gensymb}               

\def \eV{~\rm{eV}}

\def \cm{~\rm{cm}}
\def \s{~\rm{s}}

\def \g{~\rm{g}}
\def \G{~\rm{G}}

\def \erg{~\rm{erg}}

\def \yr{~\rm{yr}}

\def \keV{~\rm{keV}}

\begin{document}

\title{Common envelope jets supernovae with a black hole companion as possible high energy neutrino sources}

\author{Aldana Grichener}
\affiliation{Department of Physics, Technion, Haifa, 3200003, Israel; aldanag@campus.technion.ac.il; soker@physics.technion.ac.il}

\author[0000-0003-0375-8987]{Noam Soker}

\affiliation{Department of Physics, Technion, Haifa, 3200003, Israel; aldanag@campus.technion.ac.il; soker@physics.technion.ac.il}
\affiliation{Guangdong Technion Israel Institute of Technology, Shantou 515069, Guangdong Province, China}

\begin{abstract}
We study high energy neutrino emission from relativistic jets launched by a black hole (BH) spiraling-in inside the envelope of a red supergiant (RSG), and find that such common envelope jets supernovae (CEJSNe) are a potential source for the $\gtrsim 10^{15} \eV$ neutrinos detected by IceCube. We first use the stellar evolution code \textsc{mesa} to mimic the effect of the jets on the RSG envelope, and find that the jets substantially inflate the envelope. We then study the propagation of jets inside the extended RSG envelope and find that in most cases the jets do not penetrate the envelope but are rather stalled. We show that such jets can accelerate cosmic rays to high enough energies to produce high energy neutrinos. While the neutrinos stream out freely, the gamma-rays that accompany the neutrino production remain trapped inside the optically thick envelope. This explains the lack of observational association between high energy neutrinos and gamma-rays. We crudely estimate the diffuse neutrino spectrum from a CEJSN and find that CEJSNe with BH companions might have a substantial contribution to the high energy neutrinos flux detected by IceCube.
\end{abstract}


\section{INTRODUCTION}
\label{sec:intro}

In 2013 IceCube Neutrino Observatory detected neutrinos with energies of $E_{\rm \nu} \gtrsim 10^{15} \eV$ \citep{Aartsenetal2013}. The origin of the IceCube high energy neutrinos remains an open question. Their arrival directions are compatible with an isotropic flux (e.g., \citealt{Aartsenetal2014}), implying that there is no indication of a concentration towards either the galactic center nor the galactic plane. This strongly suggests that the high energy neutrino flux is from an extra-galactic origin, even though it does not have a significant correlation with any class of extra-galactic objects \citep{Meszaros2017}.

The production mechanism of high energy neutrinos, however, is well known. After cosmic rays are accelerated to extremely high energies by an astrophysical source, they can cool down via inelastic hadronuclear proton-proton ($\rm pp$) interactions on matter and via photohadronic ($\rm p\gamma$) interactions on photons fields  (e.g., \citealt{Stanev2006}, \citealt{DentonTamborra2018}). High energy neutrinos are emitted in the decay chain of charged pions and muons that are produced in these interactions (see section \ref{sec:NuetrinoProduction}). This emission is accompanied by the production of energetic gamma photons from the decay of neutral pions.  

Various astrophysical objects and transient events were studied as possible sources of very high energy neutrinos over the years. Among the possible candidates are blazars (e.g., \citealt{AtoyanDermer2001}; \citealt{Tavecchioetal2014}; \citealt{Oikonomouetal2019}; \citealt{LiodakisPetropoulou2020}), satuburst galaxies (e.g., \citealt{LoebWaxman2006}; \citealt{Tamborraetal2014}; \citealt{Lunardinietal2019}; \citealt{Haetal2020}), magnetars (e.g., \citealt{Iokaetal2005}) , compact binary mergers (e.g., \citealt{Xiaoetal2016}; \citealt{Kimuraetal2018}; \citealt{Decoeneetal2020} ;\citealt{Aartsenetal2020}; \citealt{Fangetal2016}; \citealt{Deyetal2016}), tidal distruption events (e.g., \citealt{LunardiniWinter2017} ;\citealt{DaiFang2017}; \citealt{Sennoetal2017}; \citealt{Steinetal2020}), X-ray binaries (e.g., \citealt{Christiansenetal2006}; \citealt{Zhangetal2010}), supermassive black holes (e.g., \citealt{Tursunovetal2020}) and gamma-ray bursts (GRBs; e.g., \citealt{WaxmanBahcall1997}; \citealt{MuraseNagataki2006}; \citealt{Niretal2016}; \citealt{Kimuraetal2017}).

The non-correlation between the IceCube high energy neutrinos and the diffuse $\gamma$-ray background measured by the Fermi gamma-ray space telescope disfavors $\gamma$-ray bright sources as the main origin of high energy neutrinos (e.g., \citealt{Aartsenetal2015}; \citealt{MuraseWaxman2016}; \citealt{Muraseetal2016}; \citealt{Aartsenetal2017}). While both fluxes are expected to be similar, the $\gamma$-ray flux is substantially lower than the neutrino flux, suggesting gamma-ray opaque sources as prime candidates for IceCube neutrinos (e.g., \citealt{Muraseetal2016}). Chocked GRB jets (e.g., \citealt{MuraseIoka2013}; \citealt{Muraseetal2016}; \citealt{Sennoetal2016}; \citealt{Fasanoetal2021}) and chocked jets in core collapse supernovae (CCSNe; e.g., \citealt{Heetal2018}; \citealt{Guettaetal2020}) have been extensively investigated as potential sources of the high energy neutrino diffuse flux. Although no events have been definitively identified as choked supernovae (SNe) jets, it is natural to suppose that some jets do not escape from the stellar envelope (e.g., \citealt{MizutaIoka2013}). It is believed that extremely energetic explosions, or explosions with trans-relativistic ejecta such as  hypernovae or Type Ic broad-line SNe, might be the result of such a scenario (e.g., \citealt{Nomotoetal2001}; \citealt{Modjazetal2016}).

In this study we suggest common envelope jets supernovae (CEJSNe) with a black hole (BH) companion as a possible source of the high energy IceCube neutrinos. 
CEJSNe are violent events powered mainly by jets that a neutron star (NS) or a BH launch as they orbit inside the envelope of a red supergiant (RSG) and then inside its core. The compact companion \footnote{Throughout the manuscript compact companion/object refers only to a NS or a BH.} accretes mass through an accretion disk, and eventually tidally destroys the core forming a massive accretion disk from the core material. The accretion disk launches two opposite jets that propagate perpendicular to the orbital plane. The collision of the energetic jets that the compact object launches with the RSG envelope converts kinetic energy to thermal energy and then radiation, leading to a bright transient event that might mimic a supernova. This is a CEJSN event. In cases where the compact object enters the envelope but does not reach the core, i.e., it removes the envelope gas before it manages to enter the core, the transient event is termed a CEJSN impostor \citep{Gilkisetal2019}. 

Earlier works found that jets in a common envelope evolution (CEE) with main sequence (MS) companions (e.g., \citealt{Shiberetal2019}) help in envelope removal. In the same manner, jets launched from a NS/BH can cause the ejection of envelope gas, therefore facilitating its removal. This can increase the common envelope efficiency above unity, i.e., $\alpha_{\rm CE} > 1$, as some studies require for CEE with compact companions (e.g. \citealt{Zevinetal2020}). However, in this work we do not take mass ejection into consideration.
     
Various studies have previously considered the mergers of NSs/BHs with cores of giant stars (e.g., \citealt{FryerWoosley1998}; \citealt{ZhangFryer2001}; \citealt{BarkovKomissarov2011}; \citealt{Thoneetal2011}; \citealt{Chevalier2012}). The CEJSN scenario differs from these merger by the key role that jets launched by the compact object play (e.g., \citealt{Papishetal2015}; \citealt{MorenoMendezetal2017}; \citealt{Moriya2018}; \citealt{Gilkisetal2019} ;\citealt{Sokeretal2019}; \citealt{GrichenerSoker2019a}; \citealt{LopezCamaraetal2019} \citealt{Schroderetal2020}). 

The violent jet-envelope interaction in CEJSNe might account for some enigmatic transient events, like the fast-rising blue optical transient AT2018cow (e.g., \citealt{Sokeretal2019}). 
\cite{SokerGilkis2018} suggest that a CEJSN can explain the peculiar supernova iPTF14hls \citep{Arcavietal2017}, in a scenario that might account for the similar SN~2020faa recently reported by \cite{Yangetal2020}. Moreover, the energetic jets of the CEJSN might serve as a possible r-process nucleosynthesis site \citep{Papishetal2015, GrichenerSoker2019a, GrichenerSoker2019b}. 

In the present work we raise the possibility that the jets launched by a BH companion in a CEJSN event can accelerate cosmic rays to sufficiently high energies to produce the IceCube high energy neutrinos ($\rm \gtrsim 10^{15} \eV$). Such jets must be relativistic. The jets of blazars, which are shown to be successful cosmic ray accelerators, are ultra-relativistic (e.g., \citealt{HovattaLindfors2019}), as are the chocked jets in failed GRBs (e.g., \citealt{DentonTamborra2018}). In X-ray binaries the jets are less relativistic with a Lorentz factor of $\rm \Gamma \approx 1.25$ and the neutrinos emitted from these jets can reach energies only up to $\approx \rm 8.5\times 10^{13} \eV$. The jets that a NS launches as it spirals-in inside the envelope of a RSG star in a CEJSN event are even slower (e.g., \citealt{LopezCamaraetal2020MN}). For jets launched in a CEJSN to be ultra-relativistic, as in the case of GRBs and collapsars, it is more likely that they are launched from a BH companion.

Our study is organized as follows. We present the CEJSN high energy neutrinos scenario in section \ref{sec:CEJSNneutrinos} and the numerical model we use to explore it in section \ref{sec:MESA}. In section \ref{sec:JetPropagationCEJSN} we study the propagation of jets in the envelope and determine the luminosities under which the jets are not able to break out. In section \ref{sec:RadiationConstraint} we show that the shocks of the jets in our scenario are not mediated by radiation, allowing for efficient cosmic rays acceleration as required for the production of high energy neutrinos. We calculate the maximum energy that the cosmic rays accelerated inside the jets launched by the BH in our scenario can reach in section \ref{sec:NuetrinoProduction}. In section \ref{sec:GammaRaysAborption} we show that in the CEJSN high energy neutrinos scenario there is no tension between the measured gamma-rays and neutrinos fluxes and in section \ref{sec:Spectrum} we crudely estimate the diffuse neutrino spectrum due to the contribution of CEJSNe. We summarize our findings and give prospects for future works in section \ref{sec:Summary}.

\section{The CEJSN high energy neutrinos scenario}
\label{sec:CEJSNneutrinos}

\begin{figure*}
\begin{center}
\includegraphics[width=0.9\textwidth]{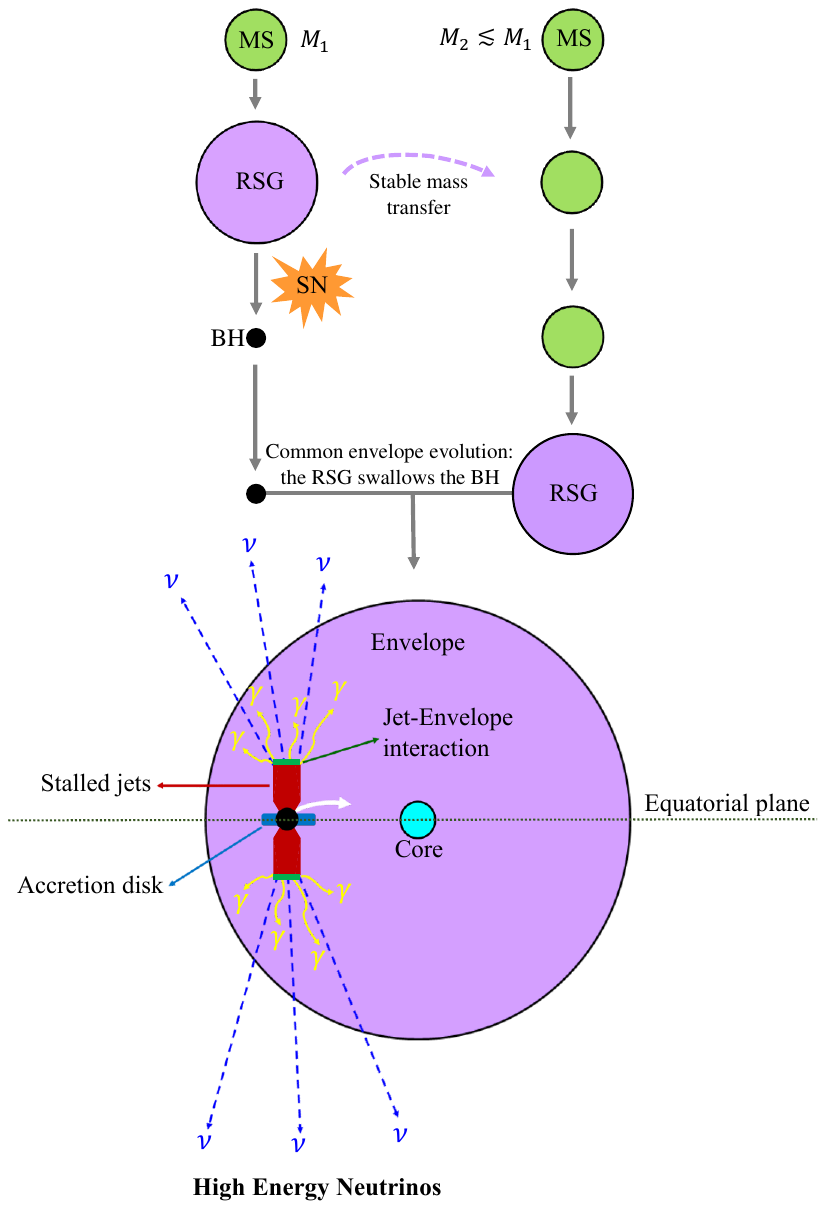}
\vspace*{-3cm}
\caption{A schematic illustration of our proposed common envelope jets supernova (CEJSN) high energy neutrinos scenario. Two massive MS stars in a binary system evolve towards a red supergiant (RSG)-black hole (BH) common envelope evolution (CEE). The BH spirals-in inside the envelope of the RSG and accretes mass through an accretion disk. The accretion disk launches relativistic jets in which energetic neutrinos and gamma-rays can be produced.}
\label{fig:CEJSN}
\end{center}
\end{figure*}

We schematically present the CEJSN high energy neutrinos scenario in Fig. \ref{fig:CEJSN}. It begins with two massive MS stars in a detached binary system. The initially more massive (primary) star of mass $\rm M_{\rm 1}$, evolves to a RSG while the less massive (secondary) star of mass $\rm M_{\rm 2}$ remains on the MS. Stable mass transfer from the primary RSG to the secondary MS star might occur, yet the MS star is massive enough to maintain synchronization and to prevent a CEE.

The RSG keeps evolving until it explodes in a CCSN. For high enough values of $\rm M_{\rm 1}$ a BH would be formed at its center. Assuming a negligible natal kick velocity, if the SN explosion ejects less than half of the binary stellar mass, the BH and the secondary star remain gravitationally bound. Later in the evolution, the secondary star evolves to become a RSG that might swallow the BH remnant of the primary star as it expands. Even if the orbital separation is up to few times the maximum radius that the RSG reaches, tidal forces would bring the BH into the envelope of the giant. 

The engulfed BH spirals-in inside the envelope of the RSG and accretes mass from the envelope, most likely through an accretion disk (e.g., \citealt{ArmitageLivio2000, Papishetal2015, SokerGilkis2018}). Studies disagree on whether an accretion disk can be formed around a stellar companion during CEE (e.g., \citealt{RasioShapiro1991, Fryeretal1996, Lombardietal2006, RickerTaam2008, Shiberetal2016, MacLeodRamirezRuiz2015a, MacLeodRamirezRuiz2015b,  MacLeodetal2017}). We adopt recent results that show the formation of an accretion disk around a compact companion inside an envelope (e.g., \citealt{Chamandyetal2018}; \citealt{LopezCamaraetal2020MN}). From simple considerations, to form an accretion disk the specific angular momentum of the accreted matter should be larger than the specific angular momentum on the companion equator. Since the radius of the last stable orbit of the BH, $\simeq 10^7 \cm$, is much smaller than the orbital separation, $\simeq 10^{9} - 10^{13} \cm$, we expect the formation of an accretion disk around the BH.  

The CEJSN scenario assumes that the accretion disk around the BH launches two relativistic jets that propagate perpendicular to the orbital plane in opposite directions. They can drill through the envelope and break out, or remain stalled inside the envelope. Whether the jets can penetrate the envelope depends on the ram pressure balance between the jets plasma and the envelope gas, which is widely affected by the orbital motion of the BH.

Relativistic jets can accelerate cosmic rays to extremely high energies (e.g., \citealt{Ostrowski2008}; \citealt{Matthewsetal2020}). In their cooling process, the cosmic rays produce energetic pions that immediately decay to very high energy neutrinos and gamma-rays. If the jets are stalled, then as the neutrinos stream out freely due to their very low cross section for interaction with matter, the gamma-rays remain trapped inside the optically thick extended envelope material, as we show in section \ref{sec:GammaRaysAborption}. This could favor CEJSNe as possible high energy neutrino emitters as they predict a much lower $\gamma$-ray flux, as observed.

\section{Numerical scheme and model settings}
\label{sec:MESA}

We use the stellar evolution code \textsc{mesa} (Modules for Experiments in Stellar Astrophysics; \citealt{Paxtonetal2010}; \citealt{Paxtonetal2015}; \citealt{Paxtonetal2018}; \citealt{Paxtonetal2019}), version 10398, to evolve the initially lighter secondary star in our proposed scenario (right column of Fig. \ref{fig:CEJSN}) from the zero age main sequence (ZAMS) to the RSG-BH CEE phase. We run a non-rotating model with a solar metalicity $\rm Z=0.02$ and an initial mass of $\rm M_{\rm 2,ZAMS}=30M_{\rm \odot}$. We follow the calculations of \cite{Schroderetal2020} and assume that a BH of mass $\rm M_{\rm BH}=3M_{\rm \odot}$ is swallowed by the giant star when it expands to a radius of $R_{\rm 2} = 1000 R_{\rm \odot}$  initiating a CEE. At this stage the mass of the secondary star in our stellar model is $\rm M_{\rm 2}=20M_{\rm \odot}$ due to mass loss through stellar winds. We further assume that the BH launches relativistic jets as it accretes mass through an accretion disk. We mimic the effect of the jets on the RSG envelope by injecting the kinetic energy of the jets into the spherically symmetric \textsc{mesa} stellar model.  

The total energy the jets deposit in the envelope is 
\begin{equation}
{E}_{\rm 2j}  = \int \eta \dot{M}_{\rm acc} c^{2} \,dt
\label{eq:Ejet}
\end{equation}
where $\eta$ is the efficiency parameter of the jets and $\dot{M}_{\rm acc}$ is the mass accretion rate onto the compact object. Typical crude values for a BH that launches jets inside the envelope of a giant star are $\eta \simeq 0.1$ (e.g., \citealt{Franketal2002}; \citealt{Schroderetal2020}) and $\dot{M}_{\rm acc}\simeq 0.1 \dot{M}_{\rm BHL}$  (e.g., \citealt{MacLeodRamirezRuiz2015b}; \citealt{Chamandyetal2018}) where $\Dot{M}_{\rm BHL}$ is the Bondi-Hoyle-Lyttleton (BHL) mass accretion rate (\citealt{HoyleLyttleton1939} ;\citealt{BondiHoyle1944}) onto the BH. For a detailed computation of $\Dot{M}_{\rm BHL}$ by a compact object inside a RSG see, e.g., \cite{GrichenerSoker2019a}.

We extract the separations between the BH and the core of the RSG from figure 1 in \cite{Schroderetal2020}, and inject the jets energy from the location of the BH outwards to the photosphere at each time interval according to equation (\ref{eq:Ejet}) with $\eta = 0.1$ and $\dot{M}_{\rm acc} = 0.1 \dot{M}_{\rm BHL}$. For each timestep the energy deposition is constant per unit mass. We continue until the BH reaches a separation of $a = 100 R_{\rm \odot}$ after 3 years, which is the numerical limit of the simulation performed by \cite{Schroderetal2020}. 

\begin{figure}
\begin{center}
\vspace*{-3.85cm}
\includegraphics[width=0.49\textwidth]{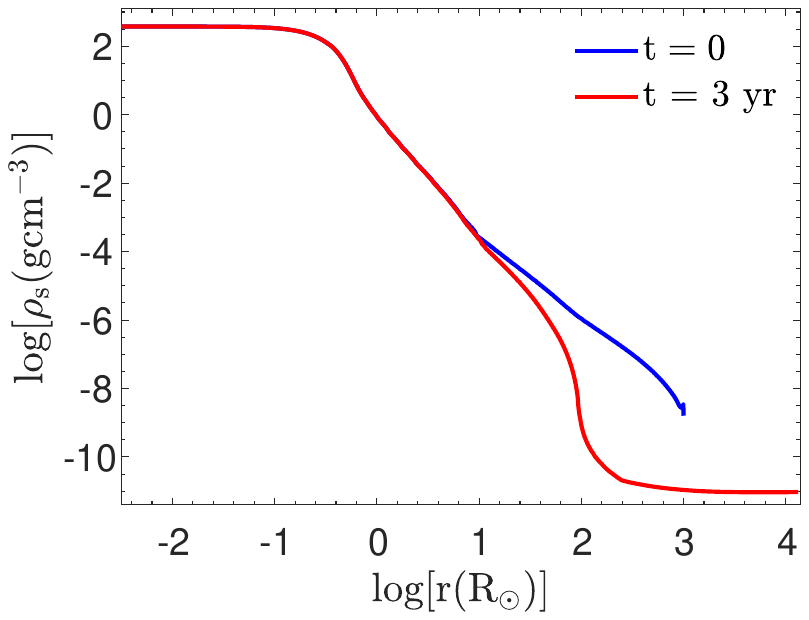}
\vspace*{-4.2cm}
\caption{Density profile of the red supergiant (RSG) when it reaches a radius of $R=1000R_{\rm \odot}$  (blue curve) and at the end of the simulation (after 3 years of energy injection) when the black hole (BH) reaches a separation of $a = 100R_{\rm \odot}$ (red curve). An extended envelope is formed around the BH-core system.}
\label{fig:DensityProfile}
\end{center}
\end{figure}

We present the density profile just before the beginning of energy injection and the density profile after 3 years in Fig. \ref{fig:DensityProfile}. We can see that the envelope gas that surrounds the BH-core binary system extends up to $R=1.3\times 10^{4} R_{\rm \odot}$. This is slightly different from the results of the $3\rm D$ simulations performed by \cite{Schroderetal2020}, whose envelope expands up to a photosphere of $R \approx 4\times 10^{4} R_{\rm \odot}$. Moreover, we note that the shape of the final density profiles in both studies is different. \cite{Schroderetal2020} obtain an extended envelope with a power-law density profile while in our case there is a jump at $R=100R_{\rm \odot}$. The differences are expected as their envelope expansion is driven by the orbital energy rather than the jets energy as in our case, and since \cite{Schroderetal2020} use a polytropic equation of state with adiabatic index $\gamma_{\rm a}=4/3$, while \textsc{MESA} has a Helmholtz equation of state.

We use the stellar density profiles from all 3 years of evolution in the subsequent derivations.

\section{Jets propagation in the extended envelope}
\label{sec:JetPropagationCEJSN}

\subsection{Jets geometry}
\label{subsec:geometry}

\begin{figure}
\begin{center}
\vspace*{-1.5cm}
\hspace*{-2.3cm}
\includegraphics[width=0.9\textwidth]{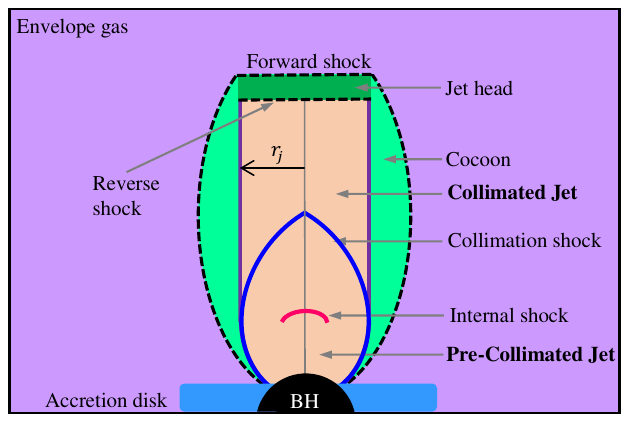}
\vspace*{-16.1cm}
\caption{A schematic illustration of a collimated jet on one side of the equatorial plane. The accretion disk around the black hole (BH) launches conical jets inside the red supergiant (RSG) envelope, which become cylindrical through collimation shocks. The strong interaction of the jet with the envelope results in a forward shock propagating into the envelope and a reverse shock facing the supersonic jet. The deep green area that separates the reverse-shocked jet and the forward-shocked envelope (dashed lines) is the jet head that expands sideways into a cocoon (light green area). Based on \cite{Brombergetal2011}. 
}
\label{fig:JetsGeometry}
\end{center}
\end{figure}

The propagation of a relativistic jet in the extended envelope is determined by its interaction with the envelope gas. As the jet drills through the envelope, a contact discontinuity is formed between the shocked jet matter and the shocked envelope gas. This region of shocked material at the front of the jet is called the \textit{jet head}. Post-shock gas with high pressure in the head region flows to the sides of the jet. The shocked jet and envelope gases form a cocoon around the jet, as we schematically draw in Fig. \ref{fig:JetsGeometry}. 

The cocoon can collimate the jet by exerting side-pressure that drives the formation of collimation shocks inside the jet (e.g., \citealt{Brombergetal2011}; \citealt{MuraseIoka2013} ;\citealt{Sennoetal2016}). These shocks converge onto the jet axis, causing the jet to maintain a constant cylindrical radius $r_{\rm j}$ above a certain height. The collimation shocks near the base of the jet have an important role in the emission of very high energy neutrinos (e.g., \citealt{MuraseIoka2013}). 

The connection between the velocity of the head ($v_{\rm h}$) and the velocity of the jet ($v_{\rm j} \simeq c$) is determined by the ram pressure balance between the shocked jet that pushes the head outwards, and the shocked envelope that decelerates the propagation of the head. This pressure balance results in (e.g., \citealt{Matzner2003})
\begin{equation}
v_{\rm h} = \frac{v_{\rm j}}{1+\Tilde{L}^{-1/2}} \simeq
\begin{cases}  c  &\mbox{if     } \Tilde{L} \gg 1 \\
\Tilde{L}^{1/2} c  \ll c   & \mbox{if     } \Tilde{L} \ll 1\end{cases}.
\label{eq:Vhead}
\end{equation}
The dimensionless parameter $\Tilde{L}$ receives different values for collimated ($\Tilde{L} < \theta_{\rm 0}^{-4/3}$) and uncollimated ($\Tilde{L} > \theta_{\rm 0}^{-4/3}$) jets (e.g., \citealt{Brombergetal2011})
\begin{equation}
\Tilde{L}\simeq 
\begin{cases}  \left(\frac{32L_{\rm 1j}}{t^{2}\theta_{\rm 0}^{4}c^{5}\rho_{\rm s}}\right)^{2/5}  &\mbox{if     } \Tilde{L} < \theta_{\rm 0}^{-4/3} \\
\frac{L_{\rm 1j}}{t^{2}\theta_{\rm 0}^{2}c^{5}\rho_{\rm s}} & \mbox{if     } \Tilde{L} > \theta_{\rm 0}^{-4/3}
\end{cases},
\label{eq:Ltilde}
\end{equation}
where $L_{\rm 1j}$ is the luminosity (due mostly to the kinetic power) of one jet, $t$ is the time since the jet was launched, $\theta_{\rm 0}$ is the initial half opening angle of the jet and $\rho_{\rm s}$ is the density of the surrounding envelope along the propagation axis. 

We use the idea that the jets operate in a negative feedback mechanism \citep{Soker2016} in which the envelope feeds the accretion disk around the BH that launches the jets. If the jets become stronger they remove more envelope mass from the vicinity of the BH, i.e., from the reservoir of the accreted gas. This reduce the mass accretion rate and consequently the jets power. The negative feedback cycle reaches an average accretion rate and therefore an average jet luminosity. Following equation (\ref{eq:Ejet}), the typical average luminosity of a jet launched by a BH that is spiralling-in in the envelope of a giant star is $L_{\rm 1j}\simeq 0.05 L_{\rm BHL}$ where $L_{\rm BHL} \equiv \eta \Dot{M}_{\rm BHL}c^{2}$. 

The main uncertainty in equation (\ref{eq:Ltilde}) is in the duration of the jet $t$. Since the BH launches the jets as it orbits inside the envelope, they start to fade after a time period that constitutes a fraction of the orbital period $P_{\rm orb}\simeq 10^{6}-10^{8} \s$ (\citealt{Papishetal2015}). 

\begin{figure}
\begin{center}
\vspace*{-5.80cm}
\hspace*{-2.0cm}
\includegraphics[width=0.69\textwidth]{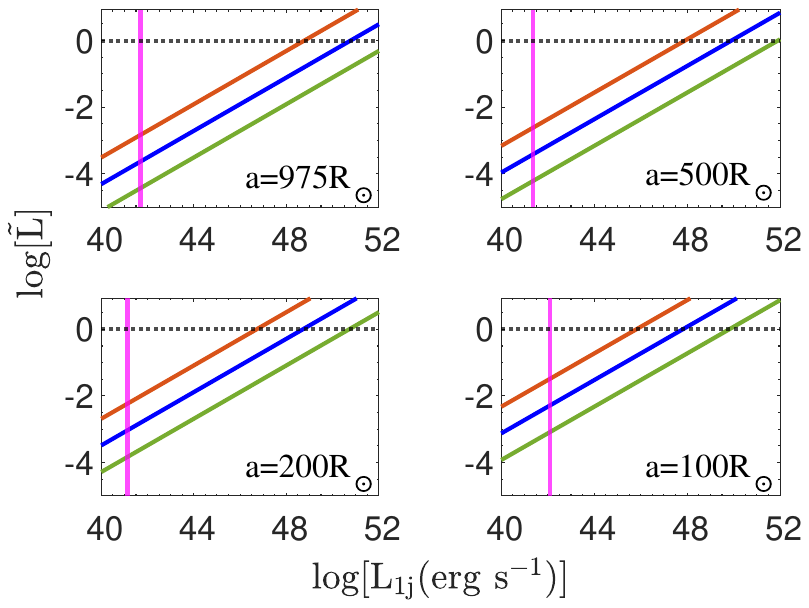}
\vspace*{-5.9cm}
\caption{The dimensionless parameter $\Tilde{L}<\theta_{\rm 0}^{-4/3}$ (equation \ref{eq:Ltilde}) for a time $t=P_{\rm orb}$ (green lower curves), $t=0.1P_{\rm orb}$ (blue middle curves) and $t=0.01P_{\rm orb}$ (brown upper curves) since the ejection of the jet as a function of the jet luminosity. The black dotted horizontal line marks $\Tilde{L}=1$. The four different panels show the results for four different orbital radii $a$. The magenta vertical line at each panel represents our estimated jet luminosity for each of the separations $L_{\rm 1j} \simeq 0.05 L_{\rm BHL}$, where $L_{\rm BHL} \equiv \eta  
\dot M_{\rm BHL} c^{2}$ and $\eta = 0.1$. We set the initial half opening angle of each jet to be $\theta_{\rm 0} = 0.2$. For the density we take the photosphere density at the point where the jet would cross the photosphere if it breaks out from the envelope, and obtain an upper bound on $\Tilde{L}$. 
}
\label{fig:CollimatedNonRelativistic}
\end{center}
\end{figure}

Fig. \ref{fig:CollimatedNonRelativistic} shows the value of $\Tilde{L}$ for a jet duration $t$ that equal different fractions of $P_{\rm orb}$ \footnote{The chosen fractions of $P_{\rm orb}$ also coincide with the range of $t_{\rm c}$ defined is subsection \ref{subsec:Stalled jets}, implying that our computations are self-consistent.}. We can see that in all cases, even for the shortest time $t=0.01 P_{\rm orb}$ (brown curve) $\Tilde{L}$ is much smaller than unity for our suggested jet luminosities. Therefore, we can safely conclude that the head velocity $v_{\rm h}$ of a jet that propagates in the envelope of our RSG star is in the non-relativistic regime. This agrees with the assumption of \cite{Heetal2018} that study the dynamics of jets in RSG envelopes. Moreover, since $\Tilde{L}\ll 1$ implies that $\Tilde{L}<\theta_{\rm 0}^{-4/3}$ , we can also conclude that the jets in our scenario are collimated as well.   

We note that the Eddington luminosity limit for the BH in our scenario, $L_{\rm Edd} = 3.5\times 10^{38} \erg \s^{-1}$, is several orders of magnitude smaller than the jet luminosities resulting from the mass accretion rates in our simulation. However, since jet ejection is associated with a non-spherical accretion, as in the case of BHL accretion, the Eddington limit does not apply and super-Eddington accretion is possible. Numerical simulations of accreting BHs have found accretion rates much higher than the Eddington limit (e.g., \citealt{Ohsugaetal2005}; \citealt{Skadowskietal2014}; \citealt{Jiangetal2014}). Moreover, neutrino cooling plays a major role in enhancing the accretion rates near compact objects by reducing the pressure near them. For a NS, neutrino cooling dominates when the mass accretion rate is $\dot M_{\rm acc} \ga 10^{-3} M_\odot \yr^{-1}$ \citep{HouckChevalier1991, Chevalier1993, Chevalier2012}. In the case of a mass-accreting BH the in-flowing gas can carry a large fraction of the energy into the BH (beyond the horizon), increasing the accretion rate needed for neutrino cooling to dominate substantially (e.g., \citealt{Pophametal1999}). For the accretion rates we find in our stellar model, the gas in-flowing into the BH carries a larger fraction of the accretion energy than the neutrinos. In any case, the accretion rate can be much higher than the Eddington limit that is calculated by assuming only photon cooling.

\subsection{Hydrodynamical constraints on stalled jets}
\label{subsec:Stalled jets}

As mentioned in subsection \ref{sec:CEJSNneutrinos}, jets that propagate inside a stellar envelope might not be able to break through (e.g., \citealt{MuraseIoka2013}; \citealt{Sennoetal2016}; \citealt{Heetal2018}). We showed in the previous subsection that the jets in our model are collimated. Therefore, the jets are unable to emerge out when their propagation time through the envelope is larger than the time it takes the BH to cross a distance that is equal to the cylindrical diameter of the collimated jet $D_{\rm j}=2r_{\rm j}$. In this case a fresh supply of jet material ceases before the jet head manages to exit the star along its propagation direction, quenching the propagation of the jet. 

As the jets propagate perpendicular to the orbital plane (along the $z$ direction), the velocity of the head is $v_{\rm h} = dz_{\rm h}/dt$. The time it takes the jets to reach the RSG surface and break out $t_{\rm bo}$ satisfies $z_{\rm h}(t_{\rm bo}) = \sqrt{R^2 - a^2}$, where $R$ is the radius of the extended envelope and $a$ is the orbital radius of the BH when it launches the jets. Substituting equations (\ref{eq:Vhead}) and (\ref{eq:Ltilde}) for a collimated jet with a non-relativistic head and performing the integration we find that
\begin{equation}
t_{\rm bo} \simeq 0.13 L_{\rm 1j}^{-1/3} \theta_{\rm 0}^{4/3} I^{5/3} \approx 10^{5} - 2\times 10^{5} \s,
\label{eq:t_bo}
\end{equation}
where $I=\int_{0}^{\sqrt{R^2 - a^2}} \rho_{\rm s}^{1/5}(z_{\rm h}) \,dz_{\rm h}$. To obtain the numerical range in equation (\ref{eq:t_bo}) we substitute the possible values of the different variables. 

Due to the significant expansion of the envelope (see Fig. \ref{fig:DensityProfile}) we can assume that no tidal synchronization has occurred, thus the BH moves with Keplerian velocity $v_{\rm K}$ with respect to the common envelope. The time it takes the BH to cross a distance equal to the cylindrical diameter of a collimated jet that reaches the tip of the envelope is then 
\begin{equation}
t_{\rm c} \simeq \frac{2r_{\rm j}(t_{\rm bo})}{v_{\rm K}} \approx 5\times 10^{3}  - 2.5\times 10^{5} \s,   
\label{eq:t_cross}
\end{equation}
where the cylindrical radius $r_{\rm j}$ is \citep{Brombergetal2011}
\begin{equation}
r_{\rm j} \simeq c^{-1/2}\theta_{\rm 0}^{4/5}L_{\rm 1j}^{3/10} t^{2/5} \bar{\rho_{\rm s}}^{-3/10} \approx 5\times 10^{11} - 10^{12} \cm, 
\label{eq:r_cylindrical}
\end{equation}
and $\bar{\rho_{\rm s}}$ is the average density along the jet axis. 

Applying the non-penetration condition $t_{\rm c}<t_{\rm bo}$ to our extended stellar model, we can find the jet luminosity $L_{\rm bo}$ for which a jet launched by the BH at a certain separation reaches the edge of the photosphere for a given initial half opening angle. Any jet whose luminosity is smaller than $L_{\rm bo}$ is not able to break out from the envelope, i.e., it remains stalled inside the envelope. 

\begin{figure}
\begin{center}
\vspace*{-4.10cm}
\includegraphics[width=0.49\textwidth]{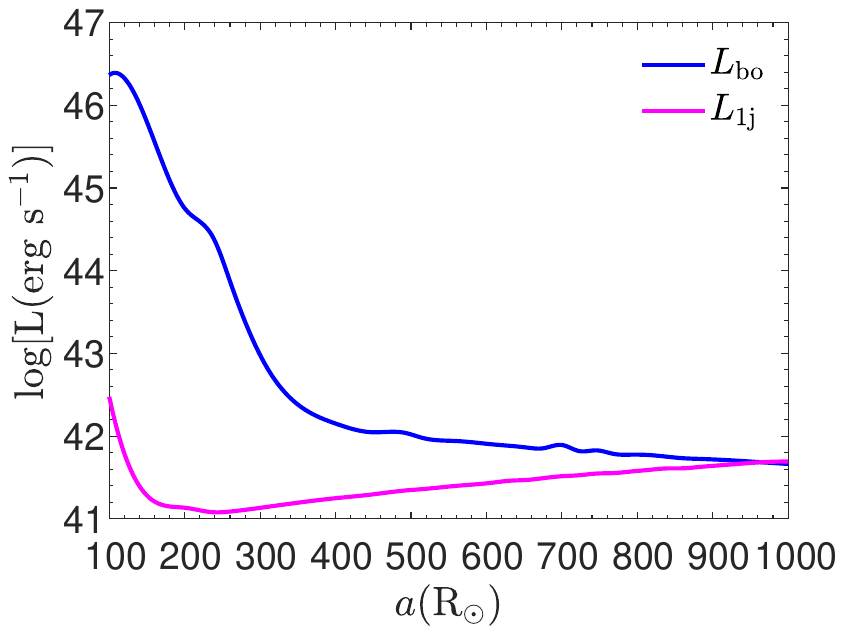}
\vspace*{-4.2cm}
\caption{Comparison between the jet luminosity in our model $L_{\rm 1j} \simeq 0.05 L_{\rm BHL}$ (magenta line) and the luminosity above which the jet would break our from the extended envelope (blue line) as a function of the orbital radius $a$ of the BH-core system. We take $\theta_{\rm 0} = 0.2$ for the initial half opening angle of the jet.
}
\label{fig:Lbreakout}
\end{center}
\end{figure}

In Fig. \ref{fig:Lbreakout} we present the breakout luminosity $L_{\rm bo}$ (blue curve) and compare it with the jet luminosity $L_{\rm 1j}$ (magenta curve) for the range of orbital radii in our simulations. We find that for jets that the BH launches at almost all orbital radii the jet luminosity in our model is lower than the value for which a jet breaks out from the envelope. This implies that the jets are stalled inside the extended envelope.

\section{Radiation constraint on particle acceleration in jets}
\label{sec:RadiationConstraint}

\subsection{Particle acceleration in shocks}
\label{subsec:TauConstraint}

Acceleration of cosmic rays can occur in shocks, such as internal shocks and collimation shocks inside jets (e.g., \citealt{MuraseIoka2013}). When a strong shock propagates, charged particles scatter back and forth across the shock interface. A proton that is scattered by the magnetic field irregularities in the upstream region of the shock would eventually find itself in the downstream region, where it can be scattered by magnetic field irregularities again. As this process repeats itself, the proton gains energy at the expense of the velocity difference between the two regions of the shock. This acceleration mechanism is called diffusive shock acceleration (e.g., \citealt{Bell1978}; \citealt{BlandfordOstriker1978}). 

Typically the timescale for scattering from magnetic field irregularities is much shorter than any timescale for particle collision in the shock, allowing for efficient cosmic rays acceleration. Such shocks are referred to as collisionless. However, it is important to take into consideration that shocks propagating inside stars might be dominated by radiation (e.g., \citealt{LevinsonBromberg2008}). In this case, photons that are produced in the downstream region of the shock diffuse into the upstream region and interact with electrons, lowering the energies of the electrons. The thermal electrons decelerate the protons. For cosmic rays to be accelerated efficiently, the upstream shock should be optically thin to electron scattering, allowing the photons to escape from this region without significantly interacting with the electrons in the flow.

Moreover, the photons that diffused into the upstream region can produce pairs of thermal electrons and positrons. These electrons interact with the protons in this region and add to the declaration of the protons. To prevent this, the upstream shock should be optically thin to pair production as well. 
 
This implies that (e.g., \citealt{MuraseIoka2013}; \citealt{Sennoetal2016}; \citealt{Heetal2018})
\begin{equation}
\tau_{\rm u} \simeq n_{\rm u}\sigma_{\rm T}l_{\rm u} \lesssim \rm min[1,0.1C^{-1}\Gamma_{\rm rel}],
\label{eq:tauAll}
\end{equation}
where $\tau_{\rm u}$, $n_{\rm u}$ and $l_{\rm u}$ are the optical depth , the proton number density (which is about equal to the electron number density) and the comoving size of the upstream region of the shock, respectively, and $\sigma_{\rm T}=6.65\times 10^{-25} \cm^{2}$ is the Thomson cross section for electron scattering. $C=1+2ln\Gamma_{\rm rel}^{2}$ is the possible effect of pair production \citep{Budniketal2010} and $\Gamma_{\rm rel}$ is the relative Lorentz factor between jet segments, which receives different values in the internal shock and collimation shock scenarios for cosmic ray acceleration (as we explain in subsections \ref{subsec:InternalShock} and \ref{subsec:CollimationShock}). The effect of pair production is especially relevant for relativistic flows where the pairs are produced in the upstream shock (e.g., \citealt{Kimuraetal2018}), although it might become small when the protons start to escape.

\subsection{Radiation constraint on particle acceleration in internal shocks}
\label{subsec:InternalShock}

Internal shocks can be formed inside a jet when a fast moving jet segment catches up with a slow segment and merges into it. The relative Lorentz factor between the two segments is $\Gamma_{\rm rel,is} \approx 3$ (e.g., \citealt{Heetal2018}). An internal shock with a height $z_{\rm is} \lesssim z_{\rm h}$ that arises in the pre-collimated zone of the jet bounds an upstream flow with a comoving size $l_{\rm u} = \frac{z_{\rm is}/\Gamma}{\Gamma_{\rm rel,is}}$  from above, where $\Gamma$ is the Lorentz factor of the jet plasma. The proton number density in the upstream region of the internal shock is $n_{\rm u}=\frac{n'_{\rm p, pc}}{\Gamma_{\rm rel,is}}$ where 
\begin{equation}
n'_{\rm p,pc} \simeq \frac{L_{\rm 1j}}{\pi \theta_{\rm 0}^{2} z_{\rm is}^{2}\Gamma^{2}m_{\rm p}c^{3}}
\label{eq:PreCollimatedRegionProtonDensity}
\end{equation} 
is the comoving proton number density in the pre-collimated jet region at the downstream of the internal shock (e.g., \citealt{MuraseIoka2013}; \citealt{Heetal2018}). 

Using all the definitions above and assuming that the internal shock arises from near the base of the jet and satisfies $z_{\rm is} < z_{\rm h} < \sqrt{R^2 -a^2}$, we can rewrite equation (\ref{eq:tauAll}) as a constraint on the jet luminosity for which the shock will not be mediated by radiation, and therefore cosmic rays acceleration in the shock will be efficient 
\begin{equation}
\begin{split}
L_{\rm 1j} \lesssim \pi m_{\rm p} c^{3} \sigma_{\rm T}^{-1} \theta_{\rm 0}^{2} z_{\rm is} \Gamma^{3} \Gamma_{\rm rel,is}^{2} \rm min[1,0.1C^{-1}\Gamma_{\rm rel,is}] \\ \simeq 
 5.3\times 10^{51} \erg \s^{-1} \left(\frac{\theta_{\rm 0}}{0.2} \right)^{2} \left(\frac{z_{\rm is}}{1000R_{\rm \odot}} \right)   \left(\frac{\Gamma}{100} \right)^{3} \\ \times \left(\frac{\Gamma_{\rm rel}}{3} \right)^{2} \rm min \left[ 1,0.06 \left(\frac{C}{5} \right)^{-1} \left(\frac{\Gamma_{\rm rel}}{3} \right) \right].
\label{eq:RadiationConstraintLis1}
\end{split}
\end{equation}

If the internal shock arises in the collimated zone of the jet the comoving size of the collimated upstream flow is $l_{\rm u} = \frac{z_{\rm is}/\Gamma_{\rm c}}{\Gamma_{\rm rel,is}}$ where $\Gamma_{\rm c}\approx \theta_{\rm 0}^{-1}$ is the constant Lorentz factor of the collimated flow (e.g., \citealt{Brombergetal2011}). The proton number density in the upstream region in this case is $n_{\rm u}=\frac{n'_{\rm p,c}}{\Gamma_{\rm rel,is}}$ where
\begin{equation}
n'_{\rm p,c} \simeq \frac{L_{\rm 1j}}{\pi \theta_{\rm 0} z_{\rm cs}^{2}\Gamma m_{\rm p}c^{3}}
\label{eq:CollimatedRegionProtonDensity}
\end{equation} 
is the comoving proton number density in the collimated jet zone (e.g., \citealt{MuraseIoka2013}). We note that even in the collimated zone the proton number density depends on the initial half opening angle $\theta_{\rm 0}$, since it determines the boundaries and the velocity of the the collimated zone. Substituting $n_{\rm u}$ and $l_{\rm u}$ into equation (\ref{eq:tauAll}) we find that for an internal shock in the collimated zone of the jet not to be dominated by radiation the jet luminosity must obey the inequality  
\begin{equation}
\begin{split}
L_{\rm 1j} \lesssim \pi m_{\rm p} c^{3} \sigma_{\rm T}^{-1} z_{\rm is}^{-1} z_{\rm cs}^{2} \Gamma  \Gamma_{\rm rel,is}^{2} \rm min[1,0.1C^{-1}\Gamma_{\rm rel,is}] \\ \simeq   
 1.3\times 10^{49} \erg \s^{-1}  \left(\frac{z_{\rm is}}{1000R_{\rm \odot}} \right)^{-1} \left(\frac{z_{\rm cs}}{1000R_{\rm \odot}} \right)^{2}   \\ \times \left(\frac{\Gamma}{100} \right)  \left(\frac{\Gamma_{\rm rel}}{3} \right)^{2}  \rm min \left[ 1,0.06 \left(\frac{C}{5} \right)^{-1} \left(\frac{\Gamma_{\rm rel}}{3} \right) \right].
\label{eq:RadiationConstraintLis2}
\end{split}
\end{equation}

As the typical jet luminosity in our model is $L_{\rm 1j}\approx 10^{41}-10^{42.5} \erg \s^{-1}$  (Fig. \ref{fig:Lbreakout}), it clearly satisfies the inequality in both equations (\ref{eq:RadiationConstraintLis1}) and (\ref{eq:RadiationConstraintLis2}). This still holds even if we take much smaller values of $z_{\rm is}$ for internal shocks closer to the BH. Therefore, we can safely conclude that the internal shocks that occur in the jets of our CEJSN scenario are not mediated by radiation, and efficient cosmic ray acceleration can take place. If the jets are launched from the core, however, the internal shocks inside the jets might be mediated by radiation due to the high accretion rates inside the core that lead to high jet luminosities.

\subsection{Radiation constraint on particle acceleration in the collimation shock}
\label{subsec:CollimationShock}

As we discussed in subsection \ref{subsec:geometry}, the jets that propagate in our extended envelope become cylindrical through a collimation shock. The collimated and pre-collimated zones of the jet are separated by a collimation shock of height $z_{\rm cs}$. The relative Lorentz factor between the two zones is $\Gamma_{\rm rel,cs}\approx 10$ (e.g., \citealt{MuraseIoka2013}). In this case, the upstream flow is the pre-collimated jet gas and its comoving size and proton density are $l_{\rm u}=z_{\rm cs}/\Gamma$ and $n_{\rm u} = n_{\rm p,cs}$, respectively, where the protons number density of the pre-collimated shock material $n_{\rm p,cs}$ is as in equation (\ref{eq:PreCollimatedRegionProtonDensity}) with $z_{is} \rightarrow z_{\rm cs}$ (e.g., \citealt{MuraseIoka2013}).

We use equation (\ref{eq:tauAll}) to find the luminosity constraint on efficient cosmic rays acceleration in the collimation shock 
\begin{equation}
\begin{split}
L_{\rm 1j} \lesssim  \pi m_{\rm p} c^{3} \sigma_{\rm T}^{-1} \theta_{\rm 0}^{2} z_{\rm cs} \Gamma^{3} \rm min[1,0.1C^{-1}\Gamma_{\rm rel,cs}] \\ \simeq  5.9\times 10^{50} \erg \s^{-1} \left(\frac{\theta_{\rm 0}}{0.2} \right)^{2} \left(\frac{z_{\rm cs}}{1000R_{\rm \odot}} \right) \\ \times  \left(\frac{\Gamma}{100} \right)^{3} 
\rm min \left[ 1,0.1 \left(\frac{C}{10} \right)^{-1} \left(\frac{\Gamma_{\rm rel}}{10} \right) \right].
\label{eq:RadiationConstraintLcs}
\end{split}
\end{equation}

Since the luminosity of a CEJSN jet that the BH launches in the extended envelope satisfies the inequality in equation (\ref{eq:RadiationConstraintLcs}), we conclude that the collimation shock is not mediated by radiation, and jets of CEJSN can efficiently accelerate cosmic rays in their collimation shock.

\section{Neutrino production inside the jets}
\label{sec:NuetrinoProduction}

\subsection{Maximum cosmic rays energy}
\label{subsec:MaxEnergy}

A likely source of high energy neutrinos should produce very high energy cosmic rays as well (e.g., \citealt{Meszaros2017}). The timescale for proton acceleration due to the magnetic field of the source in the rest frame of the jet (comoving frame) is  
\begin{equation}
t'_{\rm acc} = \frac{E'_{\rm p}}{qB'c}, 
\label{eq:t_acc}
\end{equation}
where $E'_{\rm p}$ is the comoving proton energy, $B'$ is the comoving magnetic field strength and $q$ is the proton charge in cgs units. Below variables with a prime are for the jet comoving system, while those without prime are at the rest frame of the BH or the jet head, as the head of the jet moves sub-relativisticly (subsection \ref{subsec:geometry}).

To determine the comoving magnetic field strength of the jets in our CEJSN scenario we assume that the magnetic energy density of a jet is a fraction $\epsilon_{\rm b}\approx 0.1$ (e.g., \citealt{DentonTamborra2018}) of its total energy density
\begin{equation}
\frac{B'^2}{8\pi} =  \epsilon_{\rm b} \frac{E'_{\rm 1j}(t_{\rm c})}{V'} \simeq \epsilon_{\rm b}
\frac{E_{\rm 1j}(t_{\rm c})/\Gamma}{\Gamma V},
\label{eq:MagenticEnergyDensity}
\end{equation}
where $E_{\rm 1j}(t_{\rm c})$ is the energy of a jet that is launched from a certain orbital radius and has a duration $t_{\rm c}$ (equation \ref{eq:t_cross}) and $V$ is the volume of such a jet. Assuming that the collimated jets become cylindrical through collimation shocks after a very short time, we can approximate the volume of one jet as $V \approx \pi r_{\rm j}^{2}z_{\rm s}$ where $r_{\rm j}$ is the constant cylindrical radius of the collimated jet (equation \ref{eq:r_cylindrical}), $z_{\rm s}=z_{\rm is}$ for the internal shock and $z_{\rm s}=z_{\rm cs}$ for the collimation shock. Substituting this volume into equation (\ref{eq:MagenticEnergyDensity}) and scaling with typical values we find that the comoving magnetic field strength of the jets is
\begin{equation}
\begin{split}
B' \approx 10^{3} \G  \left( \frac{\epsilon_{\rm b}}{0.1}\right)^{1/2} \left( \frac{E_{\rm 1j}(t_{\rm c})}{10^{46} \erg }\right)^{1/2} \\ \times \left( \frac{r_{\rm j}}{1.5R_{\rm \odot}}\right)^{-1} \left( \frac{z_{\rm s}}{1000R_{\rm \odot}}\right)^{-1/2} \left( \frac{\Gamma}{100}\right)^{-1} . 
\label{eq:Magenticfield}
\end{split}
\end{equation} 

Protons can be accelerated until they lose energy faster than their acceleration timescale. At high proton energies (above $10^{12} \eV$), the energy loss is mostly due to synchrotron radiation and photohadronic interactions (e.g., \citealt{Sennoetal2016}).

The typical cooling timescale due to synchrotron radiation in the comoving frame is (e.g., \citealt{DentonTamborra2018}; \citealt{Kimuraetal2018})
\begin{equation}
t'_{\rm sync} = \frac{6 \pi m_{\rm p}^{4}c^{3}}{m_{\rm e}^{2} \sigma_{\rm T} B'^{2}E'_{\rm p}},
\label{eq:t_sync}
\end{equation}
where $m_{\rm p}$ and $m_{\rm e}$ are the proton and electron mass, respectively, and $\sigma_{\rm T}=6.65\times 10^{-25} \cm^{2}$ is the Thomson cross section for electron scattering. Equating $t'_{\rm acc} \approx t'_{\rm sync}$, we find that the maximum energy a proton can reach taking into account synchrotron losses is 
\begin{equation}
E_{\rm max,sync} \approx 
 6\times 10^{20} \eV \left( \frac{B'}{10^{3} \G}\right)^{-1/2} \left(\frac{\Gamma}{100}\right),
\label{eq:Emax_sync}
\end{equation}
where $E_{\rm max,sync} = \Gamma E'_{\rm max,sync}$ is the energy in the BH frame.

Photohadronic interactions between photons and protons occur on a timescale  
\begin{equation}
t'_{\rm p\gamma} = \frac{1}{c \sigma_{\rm p\gamma} k_{\rm p\gamma} n'_{\rm \gamma}},
\label{eq:t_p_gamma}
\end{equation}
in the comoving frame, where $\sigma_{\rm p\gamma} = 10^{-28} \cm^{2}$ \citep{Muckeetal1999} is the cross section for $p\gamma$ interactions that result in multi-pion production (see subsection \ref{subsec:NeutrinoProduction}) $k_{\rm p\gamma}=0.6$ \citep{AtoyanDermer2001} is the inelasticity of the multi-pion production channel and $n'_{\rm \gamma}$ is the target photons comoving number density. We find this comoving number density $n'_{\rm \gamma} \approx  \Gamma^{2} n_{\rm \gamma}$ from the black body temperature of the thermal photons in the jet head as seen in the jet rest frame (e.g., \citealt{Sennoetal2016}) 
\begin{equation}
\begin{split}
T'_{\rm \gamma} \approx 0.35 \keV \left(\frac{\theta_{\rm 0}}{0.2} \right)^{-1/2} \left(\frac{L_{\rm 1j}}{10^{41} \erg\s^{-1}} \right)^{1/4} \\ \times \left(\frac{z_{\rm h}}{1000R_{\rm \odot}} \right)^{-1/2} \left(\frac{\Gamma}{100} \right).  
\label{eq:T_gamma_head}
\end{split}
\end{equation}
Equating $t'_{\rm acc} \approx t'_{\rm p\gamma}$, we estimate the maximum energy a proton can reach considering the energy loss from photohadronic interactions to be
\begin{equation}
E_{\rm max,p\gamma} \approx 
 4\times 10^{16} \eV  \left( \frac{B'}{10^{3} \G}\right) \left( \frac{T'_{\rm \gamma}}{0.35 \keV}\right)^{-3} \left( \frac{\Gamma}{100}\right),
\label{eq:Emax_p_gamma}
\end{equation}
where $E_{\rm max,p\gamma} = \Gamma E'_{\rm max,p\gamma}$ is the energy in the BH frame. 

We note that equation (\ref{eq:Emax_p_gamma}) is valid only above the protons threshold energy of the photo-meson production process (e.g., \citealt{MuraseNagataki2006}; \citealt{Muraseetal2014}; \citealt{Kimuraetal2018}; \citealt{Yuanetal2020}), i.e., $E'_{\rm max,p\gamma}>E'_{\rm p,th}$ where
\begin{equation}
E'_{\rm p,th} \simeq \frac{E_{\rm th} m_{\rm p} c{^2}}{E'_{\rm \gamma}},
\label{eq:t_p_th}
\end{equation}
in which $E_{\rm th}\simeq 0.15$ GeV is the threshold energy for photo-meson interactions (about the pions rest mass), $m_{\rm p} c{^2}=0.938$ GeV is the proton rest energy and $E'_{\rm \gamma} = 2.8 (T'_{\rm \gamma}/ \rm keV)\rm keV$ is the peak energy of the target photons in the comoving frame according to Wein's displacement law. Substituting the photons temperature from equation  (\ref{eq:T_gamma_head}) we find that for the parameters of our model $E'_{\rm p,th}\simeq 10^{14} \eV$. Since $E'_{\rm max,p\gamma} \simeq 4\times 10^{14} \eV$, the above condition is indeed fulfilled. Moreover, the multi-pion production channel requires $E'_{\rm max,p\gamma} \gtrsim 3E'_{\rm p,th}$ (e.g., \citealt{MuraseNagataki2006}), which is most likely satisfied for our current parameters. For smaller values of $\Gamma$, $E'_{\rm max,p\gamma}$ increases much faster than $E'_{\rm p,th}$, allowing for photo-meson production through the multi-pion channel to occur with more certainty, yet for $\Gamma\gtrsim 100$ this production channel is unlikely.

We can see that for the typical values of parameters in our scenario the greatest energy loss is due to $p\gamma$ interactions, and the characteristic maximum energy to which protons can be accelerated is given by equation (\ref{eq:Emax_p_gamma}).

Observationally, the measured value of $E_{\rm max,p\gamma}$ is lowered by the cosmological redshift $Z_{\rm red}$
\begin{equation}
E_{\rm max,p\gamma,obs} \approx \frac{E_{\rm max}}{1+Z_{\rm red}}.
\label{eq:Emax_obs_p_gamma}
\end{equation}

\subsection{Neutrino production}
\label{subsec:NeutrinoProduction}

At very high energies the dominant interactions inside the jets are between thermal photons that are produced in the jet head and propagate into the shocked regions, and the relativistic protons accelerated in the shocks (e.g., \citealt{Sennoetal2016}; \citealt{DentonTamborra2018}; \citealt{Heetal2018}). These interactions occur in the collimation shock and internal shocks of the jets. To obtain efficient cosmic rays acceleration we assume that the jets are optically thin for Thomson scattering and pair production (equation \ref{eq:tauAll}). Therefore, the photons are free to move inside the jets plasma. However, the optically thick envelope traps the photons inside the jets (see section \ref{sec:GammaRaysAborption}), facilitating their interaction with protons in the shock. The pion production efficiency can be evaluated by (e.g., \citealt{MuraseNagataki2006}) 
\begin{equation}
f_{\rm p\gamma} \simeq \rm min\left[1,\frac{t'_{\rm dyn}}{t'_{\rm p\gamma}}\right],
\label{eq:PionProductionEfficiency}
\end{equation}
where $t'_{\rm dyn} = \frac{z_{\rm s}}{c\Gamma}$. For the parameters we use in our model $t'_{\rm p\gamma} \simeq 0.04 \s$ and $t'_{\rm dyn} = 20 \s$. Therefore, we can safely conclude that in the CEJSN high energy neutrinos scenario $f_{\rm p\gamma} \simeq 1$ and almost all protons are converted to pions through the $\rm p\gamma$ process
\begin{equation}
p + \gamma \rightarrow n + \pi^{+} 
\qquad \text{and} \qquad
p + \gamma \rightarrow p + \pi^{0}, 
\label{eq:Pgamma}
\end{equation}
where $\pi$ is a pion formed in the interaction. At high energies, multi-$\pi$ production dominates and
\begin{equation}
p + \gamma \rightarrow p + a\pi^{0} + b(\pi^{+} + \pi^{-}),  
\label{eq:multiPion}
\end{equation} 
where $a$ and $b$ are integer numbers. The charged pions decay very quickly producing neutrinos and muons that decay as well
\begin{equation}
\begin{split}
\pi^{+} \rightarrow \mu^{+} + \nu_{\rm \mu}
\qquad \text{and} \qquad
\pi^{-} \rightarrow \mu^{-} + \bar{\nu}_{\rm \mu} \\
\mu^{+} \rightarrow \bar{\nu}_{\rm \mu} + e^{+} + \nu_{\rm e}
\qquad \text{and} \qquad
\mu^{-} \rightarrow \nu_{\rm \mu} + e^{-} + \bar{\nu}_{\rm e}.
\label{eq:PionMuonDecay}
\end{split}
\end{equation}
The production of high energy neutrinos is accompanied by the emission of $\gamma$-rays through the decay of neutral pions (e.g., \citealt{Stanev2006})
\begin{equation}
\pi^{0} \rightarrow \gamma + \gamma.
\label{eq:NeutralPion}
\end{equation}

From kinematic considerations, the neutrinos can carry away $\zeta \simeq 5\%$ of the protons energy (e.g., \citealt{Palladinoetal2020}).However, if the pions suffer from coolings before they decay (equation \ref{eq:PionMuonDecay}), this fraction could be much smaller (e.g., \citealt{Kimuraetal2018}). We compute the critical energies for which the cooling of pions by synchrotron, hadronic processes and adiabatic expansion (of the internal shock) becomes important, and find that for the parameters of our model these energies are several orders of magnitude larger than our typical pion energies (which are about $20\%$ of the protons energy). Performing the same computations for muons we come to the same conclusion. Therefore, meson cooling is not important in our scenario, and to produce neutrinos with observed energies $E_{\rm \nu} \gtrsim 10^{15} \eV$ as detected by IceCube, it is sufficient to accelerate protons to energies of $\approx 4\times 10^{16} \eV$ as we find in equation (\ref{eq:Emax_p_gamma}). Even if we take the more recent results of \cite{RouletVissani2020} who argue for a much smaller percentage of $\zeta \simeq1\%$, due to the large uncertainties it is still safe to conclude that the proton energies in our CEJSN high energy neutrinos scenario (equation \ref{eq:Emax_p_gamma} and \ref{eq:Emax_obs_p_gamma}) are sufficiently high to produce very energetic neutrinos.

\section{$\gamma$-ray absorption in the extended envelope}
\label{sec:GammaRaysAborption}

Emission of gamma-rays through the decay of neutral pions accompany the production of high energy neutrinos. If the gamma-rays manage to escape they will be degraded by interaction with matter in their surroundings, and the gamma-ray flux will be detected in the sub-TeV range (e.g., \citealt{Cellietal2017}; \citealt{Capanemaetal2020}). Analysis of the pion production in $\rm pp$ and $\rm p\gamma$ interactions indicates that the energy fluxes of gamma-rays and neutrinos should be similar (e.g., \citealt{Muraseetal2013}; \citealt{MuraseWaxman2016}; \citealt{Capanemaetal2020}). The isotropic diffused gamma-ray background measured by the Fermi telescope \citep{Ackermannetal2015}, however, is much lower than the neutrino flux suggested by the IceCube data. 

\cite{Muraseetal2016} find that if the IceCube neutrinos have a photohadronic origin then the sources are expected to be opaque to 1-100 GeV gamma-rays. For $E_{\rm \gamma}\simeq 1$ GeV the basic process leading to gamma-ray absorption is the formation of $e^{+}e^{-}$ pairs. In the ionized envelope gas the opacity due to pair production is \citep{GinzburgSyrovatskii1964} 
\begin{equation}
\kappa_{e^{+} e^{-}} \simeq 2.1\times 10^{-3} \cm^{2} \g^{-1} \left (ln\frac{E_{\rm \gamma}}{m_{\rm e}c^{2}} -1.9 \right),
\label{eq:OppacityPairProduction}
\end{equation}
where $E_{\rm \gamma}$ is the energy of the gamma-ray photon and $m_{\rm e}c^{2}=0.511$ MeV is the rest mass energy of an electron. The optical depth for pair production in the envelope is then
\begin{equation}
\tau_{e^{+} e^{-}}(z) = \kappa_{e^{+} e^{-}} \int_{0}^{z} \rho_{\rm s}(z') \,dz'.
\label{eq:TauPairProduction}
\end{equation}
\begin{figure}
\begin{center}
\vspace*{-4cm}
\includegraphics[width=0.49\textwidth]{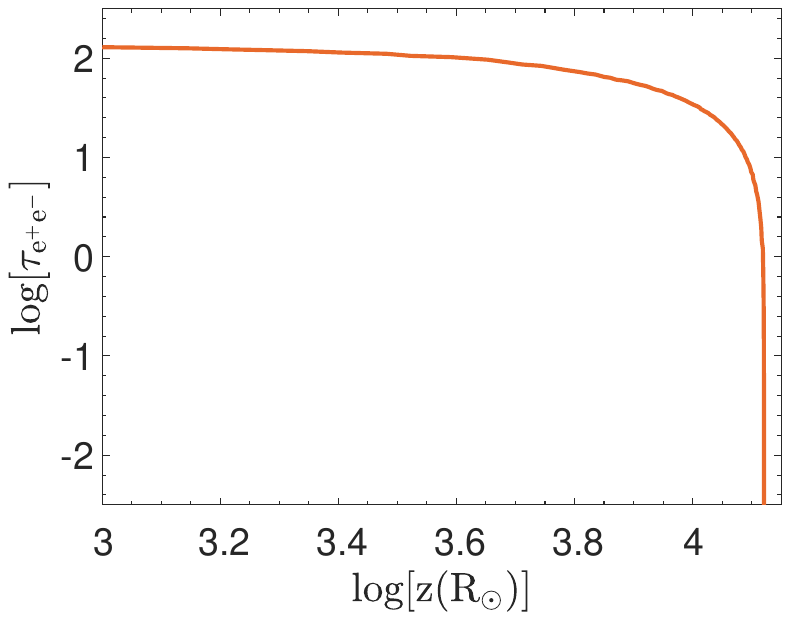}
\vspace*{-4.2cm}
\caption{The optical depth of the extended envelope due to pair production along the jet axis for gamma-rays of energy $E_{\rm \gamma}=1$ GeV. Here $z$ is the height above the equatorial plane where the gammas are emitted in the direction of the jet propagation. In this case, the jets were launched from a separation of $a = 975R_{\rm \odot}$.
}
\label{fig:OpticallyThick}
\end{center}
\end{figure}
Fig. \ref{fig:OpticallyThick} shows the optical depth of the extended envelope along the jet axis due to pair production for $E_{\rm \gamma} = 1$ GeV. Since $\tau_{\rm e^{+} e^{-}} > 10$ up to the outer layers of the extended envelope (even for jets that are launched from the largest separation we study) and for higher energies the optical depth would be even higher, we can safely conclude that CEJSN are opaque to gamma-rays in the energy range $E_{\rm \gamma} = 1-100$ GeV . Even if part of the outer extended envelope becomes unbound it will expand at a slow escape velocity remaining close to the envelope for the duration of the event and ensuring a large optical depth. 

We found earlier that for our stellar model the jets remain stalled inside the envelope if they are launched by the BH from any separation studied in our simulation. Therefore, the optically thick envelope could absorb the gamma-rays produced from the neutral pion decay, resolving the measured gamma-ray and neutrinos fluxes.

\section{Spectrum and comparison to observations}
\label{sec:Spectrum}

We crudely estimate the diffuse neutrino spectrum expected on earth due to the contribution of CEJSNe with BH companions. Previous studies compute the neutrino spectrum and the neutrino diffuse intensity from different jetted sources (e.g., \citealt{WaxmanBahcall1997}; \citealt{Heetal2012}; \citealt{Muraseetal2014}; \citealt{Yuanetal2020}). We here use the derivations of a recent study conducted by \cite{Fasanoetal2021} that computes the neutrino flux and spectrum expected from chocked GRBs. They estimate the diffuse muon neutrino flux by integrating the neutrino spectrum over different redshifts (their equation $5.3$) 
\begin{equation}
\begin{split}
E_{\rm \nu_{\rm \mu}}^{\rm obs} \frac{dN_{\rm \nu_{\rm \mu}}}{dE_{\rm \nu_{\rm \mu}}^{\rm obs}}(E_{\rm \nu_{\rm \mu}}^{\rm obs}) = \frac{c}{4\pi H_{\rm 0}} \\ \times \int_0^8 \frac{ E'_{\rm \nu_{\rm \mu}} \frac{dN_{\rm \nu_{\rm \mu}}}{dE'_{\rm \nu_{\rm \mu}}}((1+Z_{\rm red})E'_{\rm \nu_{\rm \mu}}) \frac{\Omega}{4 \pi} R(\rm Z_{\rm red}) dZ_{\rm red}}{(1+Z_{\rm red})\sqrt{\Omega_{\rm \Lambda}+\Omega_{\rm M}(1+Z_{\rm red})^{3}}},
\label{eq:DiffuseNeutrinoFlux}
\end{split}
\end{equation}
where $E'_{\rm \nu_{\rm \mu}}$ is the neutrino energy in the comoving frame, $\frac{dN_{\rm \nu_{\rm \mu}}}{dE'_{\rm \nu_{\rm \mu}}}$ is the differential neutrino spectrum at the source (in the comoving frame) which \cite{Fasanoetal2021} compute by performing Monte Carlo simulations, $\Omega = \rm 2\pi(1-cos\theta_{\rm 0})$ is the solid angle of the jet and $R(\rm Z_{\rm red})=R_0\rho(Z_{\rm red})$ is the rate of the neutrino producing event. The cosmological parameters are $H_{\rm 0}=70 \rm km s^{-1} Mpc^{-1}$, $\Omega_{\rm M}=0.3$ and $\Omega_{\rm \Lambda}=0.7$. They plot the diffuse neutrino spectrum (in units of $\rm GeV cm^{-2}s^{-1} sr^{-1}$) by taking equation \ref{eq:DiffuseNeutrinoFlux} and multiplying it by the muon neutrino energy. 

We follow the derivations of \cite{Fasanoetal2021} since we expect a similar spectrum shape for the high energy neutrinos CEJSNe scenario, and since it can be easily applied to our case. Let us analyze the similarities and differences between the parameters of equation (\ref{eq:DiffuseNeutrinoFlux}) for the case of chocked GRBs and those of the stalled CEJSN jets.

\begin{enumerate} 
\item \textit{Soild angle}. Since the half opening angle in both models is the same, $\theta_{\rm 0}=0.2$, the jets in both cases have the same solid angle.
\item \textit{Differential neutrino spectrum at the source}. A fraction of the source energy is converted to high energy neutrinos. Therefore, we can crudely scale the differential neutrino spectra of chocked GRBs and CEJSN by the ratio of jets energy in both events. \cite{Fasanoetal2021} take the energy of each chocked GRB jet to be $10^{51} \erg$. In our case the average energy of a jet that operates continuously throughout the CEJSN high energy neutrino scenario is $\approx 10^{49} \erg$. Therefore we scale the differential neutrino spectrum by a factor of $10^{-2}$ with respect to that of \cite{Fasanoetal2021}. 
\item \textit{Rate}. \cite{Fasanoetal2021} assume that the rate of chocked GRB jets at redshift $Z_{\rm red}$ is $R(Z_{\rm red})=R_{\rm 0} \rho(Z_{\rm red})$ where $\rho(Z_{\rm red})$ is the unit-less relative star formation rate and $R_{\rm 0}$ is the intrinsic local rate of chocked GRBs in units of $\rm Gpc^{-3}\yr^{-1}$. The effect of beaming is accounted for by the factor $\frac{\Omega}{4\pi}$. Since the time that elapses between star formation and CEJSNe is short compared to the typical timescale of star formation itself, we can also safely assume that the rate of CEJSNe follows the star formation rate. The difference in the event rate lies in the local rate of CEJSN high-energy neutrino sources $R_{\rm 0,CEJSN}$, to that of chocked GRBs, $R_{\rm 0,Chocked}$.   
\end{enumerate}

The local rate of CCSNe is estimated to be $\approx 1 \times 10^{5} \rm Gpc^{-3}\yr^{-1}$ (e.g., \citealt{Lietal2011}). The estimated rate of BH-core mergers is about $0.3\%$ of the rate of CCSNe \citep{Schroderetal2020}. This implies that the local rate of CEJSNe with BH companions according to population synthesis is about $R_{\rm 0,CEJSN,ps} = 300 \rm Gpc^{-3}\yr^{-1}$. For chocked GRBs, \cite{Fasanoetal2021} compare the predicted flux of muon neutrinos on earth with the results obtained by IceCube and find that the best agreement between the data and their simulation occurs for a local rate of $R_{\rm 0,Chocked} \approx 1 \rm Gpc^{-3}\yr^{-1}$, when they assume that almost the entire energy of the jets goes to accelerate protons. If this is not the case, the event rate required to explain IceCube observations would be higher. Due to the energy differences between both scenarios, to coincide with observations the local rate of CEJSNe that produce high energy neutrinos should be $100$ times higher, i.e., $R_{\rm 0,CEJSN,\nu} \approx 100 \rm Gpc^{-3}\yr^{-1}$, which is three times lower than the local rate we estimated above. The inequality $R_{\rm 0,CEJSN,\nu} < R_{\rm 0,CEJSN,ps}$ does not constitute a problem due to the vast uncertainties regarding the energies of the jets and the fraction of jets energy that goes to proton acceleration in both scenarios. 

A new publication of the IceCube collaboration (\citealt{Aartsenetal2020b}; Fig. 10) lists the required local event rate as function of the total neutrino energy per event. If we assume that about a third of the jets energy escapes as energetic neutrinos, then the total neutrino energy in our scenario is $\approx 3 \times 10^{48} \erg$ for the jets energy that we estimate here. In this case the event rate of CEJSNe with BH companions that is required to account for all IceCube high energy neutrinos should be $R_{\rm 0,CEJSN,\nu} \approx 3000 \rm Gpc^{-3}\yr^{-1}$, which is an order of magnitude higher than $R_{\rm 0,CEJSN,ps}$. For a case where a larger fraction of the jets energy results in energetic neutrinos, the required local event rate can be much lower.

We note that there are two main factors that can affect the jets energy in our scenario. First, the jets operate in a negative feedback mechanism, implying that they have episodes of quiescence throughout the CEJSN event. These episodes of very low activity would reduce the time-average energy of the jets, and therefore increase the required local rate in order to account for the IceCube neutrino flux (according to equation \ref{eq:DiffuseNeutrinoFlux} and Fig. 10 in \citealt{Aartsenetal2020b}). On the other hand, we followed the spiraling-in of the BH in our simulation only until it reached an orbital separation of $a=100R_{\rm \odot}$ after three years. Deep inside the envelope the luminosity of the jets would be much higher due to higher density of the surrounding gas. Therefore, the actual energy of the jets might be substantially larger than the energy we estimated in this manuscript, leading to a lower required event rate in order to explain IceCube observations. For jets energy of $\approx 10^{50} \erg$, i.e., that are about an order of magnitude higher than in our case, the required local event rate would be $R_{\rm 0,CEJSN,\nu} \approx 500 \rm Gpc^{-3}\yr^{-1}$. This implies that our proposed CEJSN high energy neutrino source has the potential to substantially contribute to the Ice Cube high-energy neutrino flux.

\section{Summary and prospects of future work}
\label{sec:Summary}

We examined the possibility that the relativistic jets that a BH launches as it spirals-in inside the envelope of a RSG emit high energy neutrinos. We found that in our new proposed scenario the two opposite relativistic jets accelerate cosmic rays to high enough energies to produce neutrinos with energies up to $E_\nu \gtrsim 10^{15} \eV$, as reported by IceCube \citep{Aartsenetal2013}.   
We described the evolution of the CEJSN high energy neutrino scenario in section \ref{sec:CEJSNneutrinos} (Fig. \ref{fig:CEJSN}). We assumed that the jets deposit their  energy into the envelope of the RSG star (equation \ref{eq:Ejet}), and we used the 1D stellar evolutionary code \textsc{mesa} to mimic the effect of the jets on the envelope (section \ref{sec:MESA}), mainly in inflating the envelope (Fig. \ref{fig:DensityProfile}).

Using the numerical profiles of the extended RSG envelope, we studied the propagation of the relativistic jets (section \ref{sec:JetPropagationCEJSN}) and found that the jet head is not relativistic, and that an initially conical jet becomes cylindrical through a collimation shock. We compared the time it takes such jets to break out from the envelope (equation \ref{eq:t_bo}) with the time in which the BH crosses a distance that is equal to the cylindrical diameter of the collimated jet (equation \ref{eq:t_cross}), namely the time that the BH supplies fresh jet material to any specific cross section.  We found that for our typical jet luminosities, in most cases the jets remain stalled inside the extended envelope (Fig. \ref{fig:Lbreakout}). Therefore, the optically thick envelope traps the gamma-rays that the jets produce alongside the high energy neutrinos (section \ref{sec:GammaRaysAborption}). 

The internal and collimation shocks of the jets, however, are optically thin to electron scattering and pair production for any reasonable jet luminosity, allowing for efficient cosmic rays acceleration inside both shocks (section \ref{sec:RadiationConstraint}). In section \ref{sec:NuetrinoProduction} we made a crude estimate of the maximum proton energy the jets in our scenario can achieve through acceleration (equation \ref{eq:Emax_p_gamma}) and found that it is sufficiently high to produce neutrinos up to energies of $E_{\rm \nu} \gtrsim 10^{15}$. 
We expect the neutrino spectrum of our proposed scenario to be similar to that from chocked GRBs. For that, we adopted the results of \cite{Fasanoetal2021} for chocked GRBs (here equation \ref{eq:DiffuseNeutrinoFlux}) yet we note that the jets in our scenario are much weaker than their choked GRB jets. This results in different local rates in both cases. To account for the observed high-energy neutrino flux given the jets energy we estimated for a BH spiralling-in in the outer envelope, the local CEJSN high neutrino rate is required to be $R_{\rm 0,CEJSN,\nu} \approx 3000 \rm Gpc^{-3}\yr^{-1}$. Taking into account that the energy of the jets might be much higher since the BH continues to spiral-in deep into the dense envelope of the RSG, the required local event rate is much lower, and could be compatible with $R_{\rm 0,CEJSN,ps}$.
Overall, our main conclusion is that CEJSNe with a BH companion might be a promising source of the high energy neutrinos, although probably not the only one.  

Further investigation of CEJSNe as high energy neutrino emitters is required to derive more realistic proton and neutrino energies. To find a more accurate neutrino spectrum of a single BH-CEJSNe event, future calculations should take into account all the possible cooling processes of cosmic rays, as well as the detailed microphysics of pion production and decay. Moreover, we should keep in mind that high density environments can affect the measured neutrino spectrum by leading to unique features of neutrino oscillation \citep{CarpioMurase2020}. To determine the contribution of CEJSNe to the diffuse high energy neutrino flux it is important to follow the spiraling-in of the BH deeper inside the envelope and find more accurately the total energy of the jets in a CEJSN event. 3D hydrodynamical simulations of CEJSN with a BH companion would also improve the accuracy of our results. These will be the topics of our future works in the field.

\acknowledgments

We thank Dafne Guetta, Diego L{\'o}pez-C{\'a}mara, Hila Glanz, Kohta Murase and Sophie Lund Schr{\o}der for helpful comments. We thank an anonymous referee for correcting errors in the first version of the manuscript and for very detailed comments that helped in improving our paper. This research was supported by a grant from the Israel Science Foundation (769/20). A.G. acknowledges support from the Excellence Scholarship of Prof. Pinchy Foundation.

\textbf{Data availability: The data underlying this article will be shared on reasonable request to the corresponding author.}

\end{document}